\documentclass[a4paper,11pt,dvipsnames]{article}

\usepackage[T1]{fontenc}
\usepackage[utf8]{inputenc}
\usepackage[english]{babel}

\usepackage{amsthm,mathtools}
\usepackage{thmtools,thm-restate}

\usepackage[charter]{mathdesign}
\usepackage[varqu,varl]{inconsolata}

\usepackage[margin=3cm]{geometry}
\usepackage{csquotes,microtype,xcolor}
\usepackage{graphicx,siunitx,setspace}

\usepackage[colorlinks,unicode,bookmarksnumbered]{hyperref}
\usepackage[nameinlink,capitalise,noabbrev]{cleveref}
\usepackage{bookmark}

\usepackage{caption}
\hypersetup{linkcolor=RoyalBlue, citecolor=MidnightBlue, urlcolor=RoyalBlue}
\hypersetup{pdfauthor={Ulysse Herbach},pdftitle={Gene regulatory network inference from single-cell data using a self-consistent proteomic field},pdfsubject={Preprint}}

\usepackage[pagestyles]{titlesec}

\newpagestyle{main}{
\sethead[\ifthesection{\em\thesection\hspace{0.5em}\sectiontitle}{\em\sectiontitle}][][]
{\ifthesection{\em\thesection\hspace{0.5em}\sectiontitle}{\em\sectiontitle}}{}{}
\headrule\setheadrule{0.5pt}\setfoot{}{\thepage}{}}

\renewpagestyle{plain}{\setfoot{}{\thepage}{}}

\newcommand*{\R}{\ensuremath \mathbb{R}} 
\newcommand*{\intd}[1]{\ensuremath \mathrm{d}{#1}} 
\newcommand*{\dr}{\ensuremath \partial} 
\newcommand*{\Poi}{\ensuremath \mathrm{Poisson}} 
\newcommand*{\Gam}{\ensuremath \mathrm{Gamma}} 
\newcommand*{\Bet}{\ensuremath \mathrm{Beta}} 
\newcommand*{\konb}{\ensuremath k_{\mathrm{on}}}
\newcommand*{\koffb}{\ensuremath k_{\mathrm{off}}}
\newcommand*{\kon}[1]{\ensuremath k_{\mathrm{on},#1}}
\newcommand*{\koff}[1]{\ensuremath k_{\mathrm{off},#1}}

\crefname{equation}{}{}

\title{\vspace{-15mm}\textbf{Gene regulatory network inference from single-cell data using a self-consistent proteomic field}}

\author{Ulysse Herbach\thanks{Université de Lorraine, CNRS, Inria, IECL, F-54000 Nancy, France (\href{mailto:ulysse.herbach@inria.fr}{ulysse.herbach@inria.fr})}}

\date{}

\begin{document}

\thispagestyle{plain}
\pdfbookmark[section]{Title}{title}

\maketitle

\vspace{-5mm}

\begin{abstract}
The well-known issue of reconstructing regulatory networks from gene expression measurements has been somewhat disrupted by the emergence and rapid development of single-cell data. Indeed, the traditional way of seeing a gene regulatory network as a deterministic system affected by small noise is being challenged by the highly stochastic, bursty nature of gene expression revealed at single-cell level. In previous work, we described a promising strategy in which network inference is seen as a calibration procedure for a mechanistic model driven by transcriptional bursting: this model inherently captures the typical variability of single-cell data without requiring ad hoc external noise, unlike ordinary or even stochastic differential equations often used in this context. The resulting algorithm, based on approximate resolution of the related master equation using a self-consistent field, was derived in detail but only applied as a proof of concept to simulated two-gene networks. Here we derive a simplified version of the algorithm and apply it, in more relevant situations, to both simulated and real single-cell RNA-Seq data. We point out three interesting features of this approach: it is computationally tractable with realistic numbers of cells and genes, it provides inferred networks with biological interpretability, and the underlying mechanistic model allows testable predictions to be made. A practical implementation of the inference procedure, together with an efficient stochastic simulation algorithm for the model, is available as a Python package.
\end{abstract}

\section{Introduction}

Gene regulatory networks, which typically arise from interactions between genes through protein production, are at the core of many aspects of cellular behavior such as differentiation: their automated reconstruction from large datasets has become a classic task in systems biology~\cite{Huynh-Thu2019}. Indeed, even though it may not be possible to reconstruct all interactions perfectly, there is a clear interest in statistical tools to assist experiments, for example by revealing particular patterns in a network or by making predictions about the behavior of a cell after acting on a particular gene.

\subsection{Gene expression data}

The most accessible data at present are mRNA levels, also known as expression measurements or transcriptomic data, which can be broadly divided into two types, \emph{population} and \emph{single-cell}. The second type contains the richest information: at a given time point, one simultaneously measures expression levels of a given set of genes in a number of individual cells. From a statistical viewpoint, we therefore have access to a \emph{joint probability distribution} of gene expression, while population data simply correspond to the average of each gene. An important detail is that measurement techniques usually involve the physical destruction of the cell: when measurements are made at several time points, for example to study convergence to a possibly new steady state after applying a perturbation, one does not obtain cell trajectories but rather independent samples of the time-varying distribution, sometimes referred to as \emph{snapshots} or \emph{cross-sectional profiles}~\cite{Ocone2015,Richard2016,Papili-Gao2018}.

\begin{figure}[t]
\includegraphics[width=\textwidth]{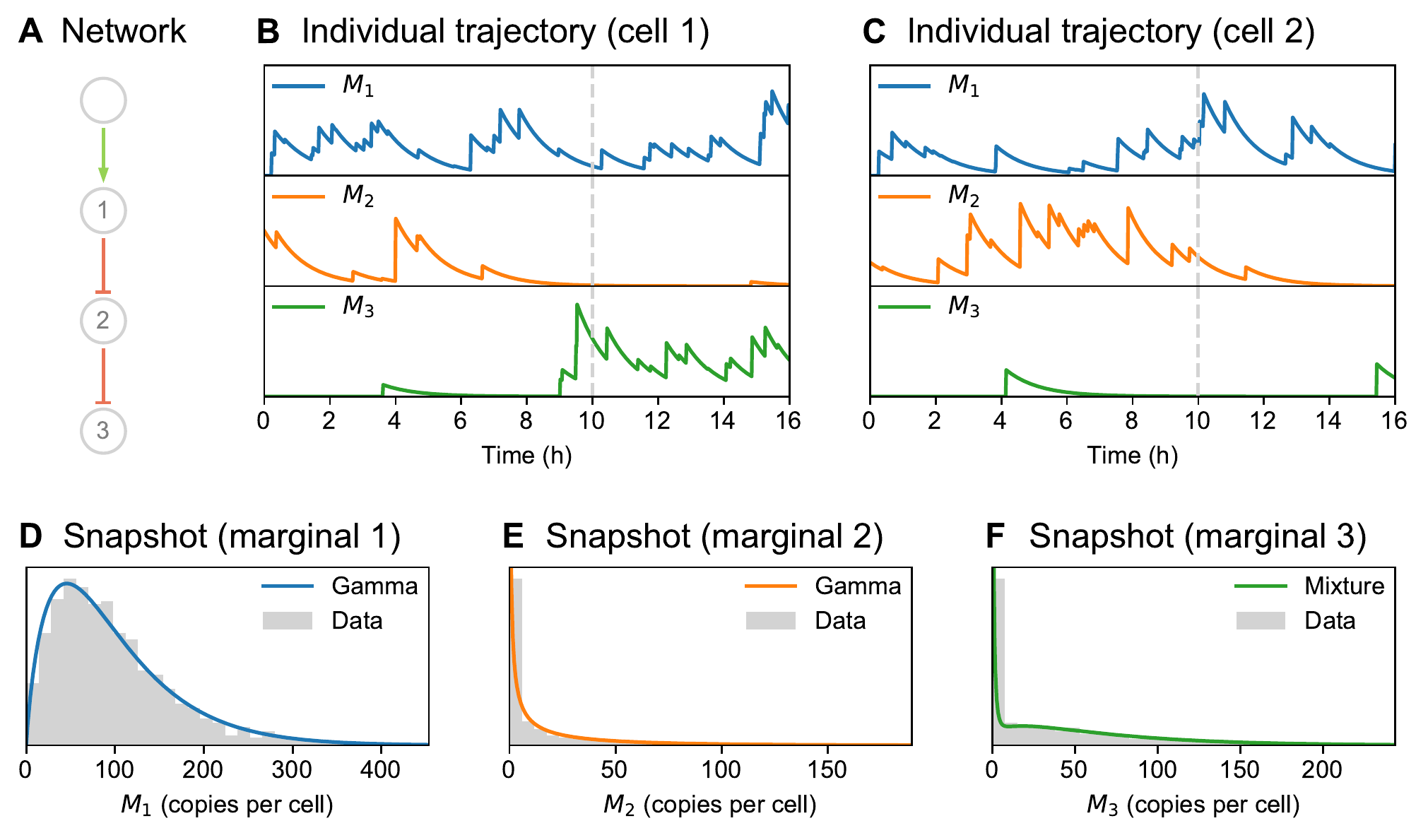}
\caption{Dealing with gene expression in individual cells. (A)~Example of 3-gene network with a stimulus (unnumbered node). Activation and inhibition are shown in green and red, respectively. (B)~Corresponding cell trajectory (with mRNA levels of genes $1$, $2$ and $3$ being denoted by $M_1$, $M_2$ and $M_3$), as obtained by simulating the mechanistic model considered in this paper. The dashed gray line indicates what would be observed in practice, i.e., a time-stamped snapshot. (C)~Another cell trajectory from the same model. While network logic turns out to be accurately reproduced at the statistical level, such cell-to-cell variability is expected to be crucial in the context of gene network reconstruction. (D)~Marginal distribution of gene~$1$ from simulated snapshot data (2000 cells) obtained at $t=\SI{10}{h}$, well approximated by a gamma distribution in this case. (E)~Marginal distribution of gene~$2$, also well approximated by a gamma distribution with different parameters than gene~$1$. (F)~Marginal distribution of gene~$3$, better approximated by a mixture of two gamma distributions combining the previous cases.}
\label{fig1}
\end{figure}

Such datasets are now widely accessible and the main challenge for biologists is to find relevant tools to analyze them~\cite{Luecken2019}. In practice, a wide range of multivariate methods have been applied to expression data and more specifically to gene network inference~\cite{Huynh-Thu2019,Hecker2009,Wang2014,Nguyen2020}. For example, one popular inference algorithm is based on a regression approach using random forests~\cite{Huynh-Thu2010,Aibar2017} whereas principal component analysis (PCA) remains an essential first tool~\cite{Luecken2019}. Interestingly, while population data are typically in good accordance with the normal distribution paradigm, single-cell data are often better described by gamma distributions (\cref{fig1}) or multimodal mixtures of gamma distributions~\cite{Mar2019,Chen2018}. The latter can sometimes be successfully simplified by basic binarization using a threshold~\cite{Moignard2015}, and from a mechanistic perspective it becomes increasingly clear that such non-normality is closely related to the well-acknowledged bursty nature of gene expression~\cite{Bahar-Halpern2015,Smirnov2018,Herbach2017}.

\subsection{Current challenges}

Despite the large number of computational tools currently available to reconstruct gene regulatory networks from expression measurements, it must be admitted that the issue can still hardly be described as solved.
As enlightened by recent comprehensive investigations, some well-established methods based on population data seem to perform poorly on single-cell data~\cite{Chen2018}, and even recent algorithms specially designed for single-cell data tend to have limited accuracy~\cite{Pratapa2020,Qiu2020}.
Overall, it can be very straightforward to generate directed or undirected graphs from data (e.g., applying a threshold to correlation coefficients), but very difficult to assess whether these graphs correspond to some biological reality.

More specifically, many existing methods focus on statistical patterns in data without considering a mechanistic description of gene expression, making biological interpretation of inferred graphs less straightforward~\cite{Huynh-Thu2010,Gallopin2013,Ghazanfar2016,Chan2017}. A basic but instructive example is given by so-called \emph{correlation networks}, which only describe pairwise gene correlations~\cite{Huynh-Thu2019}: within this framework, if a gene activates a second gene which in turn activates a third one, the logically inferred link between the first and third genes has a true statistical interpretation but no biological meaning. Our interest, which corresponds in this example to the notion of \emph{partial correlation}, is precisely to avoid such nonbiological links.

In addition, since different modeling hypotheses may lead to different statistical features, the concept of \enquote{realistic benchmark} is itself unclear. We argue that it should essentially follow current biological knowledge, which has been rapidly progressing with the development of measurements in individual cells. For instance, network models should at least be able to generate gamma distributions when describing continuous data~\cite{Albayrak2016} and gamma-mixtures of Poisson distributions (also known as negative binomial distributions) when describing discrete data~\cite{Singer2014}, since these patterns are now acknowledged to arise biologically from the transcriptional bursting phenomenon~\cite{Shahrezaei2008,Friedman2006}.

\subsection{Mechanistic approach}

In this context, it may be fruitful to consider a mechanistic approach to gene regulatory network inference, that is, directly related to a biochemical model. This idea is not new since various ODE-based dynamical models have already been used to describe population data~\cite{Mizeranschi2015}, which we now interpret as average trajectories, but there is still no consensual model of such average trajectories in terms of biochemical processes. Besides, single-cell data provide two important features related to cell-to-cell variability: (1) access to statistical dependencies between genes, which had only been accessible indirectly by perturbing gene expression (e.g., knockouts), potentially offers a huge gain in accuracy of network observation; (2) access to the distribution, and not only its mean, may allow deeper modeling and thereby a linear description at microscopic level, even if the dynamics becomes nonlinear at macroscopic level.

After being considered for some time as “noise” for cells, this variability has become well accepted in literature~\cite{Kaern2005} and it is now clear that it plays an important and sometimes biologically relevant role~\cite{Huang2009,Eldar2010,Moris2018,Guillemin2019}. As also observed in~\cite{Singer2014} and exploited in~\cite{Gu2015}, multimodal cases are well described by mixtures of the above gamma distributions, which may correspond to different \emph{burst frequencies} and reflect metastable states emerging from interactions between genes~\cite{Herbach2017,Ventre2020}. In this view, the notion of cell state---and even that of cell type---directly relates to this metastability~\cite{Mojtahedi2016,Moussy2017}.

In this paper, we consider a mechanistic network model that is able to reproduce observations, this aspect being seen as \emph{model simulation}, while also providing an effective inference algorithm, this aspect being seen as \emph{model calibration}. The network model and the inference procedure were implemented together as a Python package called HARISSA (HARtree approximation for Inference along with a Stochastic Simulation Algorithm) available at \url{https://github.com/ulysseherbach/harissa}. Although focusing on underlying ideas rather than practical implementation, the present paper describes both modeling and inference parts of the HARISSA package in \cref{sec_mechanistic_model} and \cref{sec_inference_procedure}, respectively.

We still need to clarify what is meant by “network” here. We shall make a clear distinction between fundamental interactions between genes, which emerge from their physical structure and will be assumed fixed at our time scale (development of a single organism), and statistical dependencies between expression levels, which are likely to vary over time or cell states.
Hence, a network will have a static \emph{structure} representing a given set of interactions by a directed graph, while the joint distribution of expression levels at a given time point---the co-expression probabilistic pattern---will correspond to possibly dynamic \emph{states} of the network.

A simple example is the activation of a signaling pathway: a gene that was previously not expressed starts to be transcribed, then mRNA is translated into proteins that in turn trigger transcription of another gene, and so on. For us, this pathway is always present in the network, but it is simply not detectable until the first gene reaches a certain level of expression. From the inference viewpoint, it is clear that we will only be able to reconstruct this pathway if our observations correspond to a period when it is used: to make a simplistic analogy, one cannot know which bulbs are plugged in when there is a power failure. The aim is therefore not to have the \enquote{complete network}, which would correspond to an exhaustive knowledge of the cell's functioning, but rather to reconstruct parts of this network associated with a particular phenomenon such as differentiation.

Finally, we emphasize that our aim is to use time-stamped snapshot data directly, without resorting to any preliminary inference of putative cell trajectories (i.e., pseudotime ordering of cells) which might not be reliable in the context of cellular decision making~\cite{Moris2016}.
We shall rather exploit real, experimental time points and embrace cell-to-cell variability using the notion of (possibly time-varying) quasipotential energy landscape~\cite{Huang2009,Mojtahedi2016,Moris2016}, where cells statistically tend to occupy lower energy regions of the high-dimensional gene expression space: in practice, our approach can be seen as a way to infer such a landscape from data.
Besides, we strive to deal not only with common single-cell RNA-Seq, but also with more accurate techniques such as RT-qPCR~\cite{Richard2016,Mojtahedi2016,Moussy2017} or RT-ddPCR~\cite{Albayrak2016}. The question of RNA-Seq specific limitations is not central in this work but will be discussed in \cref{sec_results} (Results) and \cref{sec_discussion} (Discussion and prospects).

\section{Mechanistic model}\label{sec_mechanistic_model}

In this section, we briefly introduce a mechanistic network model detailed in previous work~\cite{Herbach2017}. We then consider its bursty regime to obtain a simplified version, which will be our standard for simulating networks.

\subsection{Promoter-based network}

Following~\cite{Herbach2017}, our starting point is a stochastic model describing mRNA and protein levels produced by a set of $n$ genes with two-state promoters.
Mathematically, it can be interpreted as a piecewise-deterministic Markov process (PDMP) corresponding to a system of randomly switching ordinary differential equations. This stochastic process consists of three components for each gene $i=1,\dots,n$: the promoter state $E_i=0$ or $1$ (inactive or active), the mRNA level $M_i$ and the protein level $P_i$.
To simplify notation, we shall write $E = (E_1,\dots,E_n)$, $M = (M_1,\dots,M_n)$ and $P = (P_1,\dots,P_n)$.
The dynamics of each gene is then given by the following ordinary differential equations:
\begin{align}
\dot{M_i} &= s_{0,i} {E_i} - d_{0,i} {M_i} \label{eq_full_m} \\
\dot{P_i} &= s_{1,i} {M_i} - d_{1,i} {P_i} \label{eq_full_p}
\end{align}
with additional specification that $E_i$ can randomly switch from $0$ to $1$ (activation) with protein-dependent rate $\kon{i}(P)$, and from $1$ to $0$ (inactivation) with constant rate $\koff{i}$.
Biologically, \cref{eq_full_m} corresponds to transcription at rate $s_{0,i} {E_i}$ and mRNA degradation at rate $d_{0,i}$, while \cref{eq_full_p} denotes translation at rate $s_{1,i} {M_i}$ and protein degradation at rate $d_{1,i}$.
Hence, interactions between genes emerge from promoter activation rates which in turn control mRNA and protein levels.

When $\kon{i}$ is constant, this PDMP model can be derived as a rigorous hybrid approximation of the related discrete-value chemical reaction system, while being much faster to simulate (see~\cite{Ball2006,Kang2013} for a general methodology and~\cite{Crudu2012} for a theorem-based presentation). In practice, its accuracy remains satisfactory in the general case~\cite{Lin2016,Herbach2017}.
Note that $\kon{i}$ can be virtually any function of $P$ implementing dependence of gene $i$ on other genes: we refer to~\cite{Herbach2017} for biological and mathematical details. Basically, gene $i$ is said to \emph{activate} (resp. \emph{inhibit}) gene $j$ when $\kon{j}$ is an increasing (resp. decreasing) function of protein $P_i$.
To describe the system evolving after some initial perturbation, we consider the possibility of a stimulus symbolized by gene $0$, with $P_0(t)=0$ for $t \leqslant 0$ and $P_0(t)=1$ for $t > 0$. An example with three genes plus stimulus is given in \cref{fig1}.

In order to both reproduce real data and perform inference, we define
\begin{equation}\label{eq_kon_base}
\kon{i}(P) = \frac{k_{0,i} + k_{1,i} \exp(\beta_i + \sum_{j} \theta_{ji} P_j )}{1 + \exp(\beta_i + \sum_{j} \theta_{ji} P_j )}
\end{equation}
which can be rewritten as $\kon{i}(P) = (1-\sigma_i(P))k_{0,i} + \sigma_i(P)k_{1,i}$ where
\begin{equation}\label{eq_sigma_base}
\sigma_i(P) = \frac{\exp(\beta_i + \sum_{j} \theta_{ji} P_j )}{1 + \exp(\beta_i + \sum_{j} \theta_{ji} P_j )}
\end{equation}
such that the promoter activation rate is comprised between $k_{0,i}$ and $k_{1,i}$. The parameter $\beta_i$ represents basal activity of gene $i$, and $\theta_{ji}$ encodes interaction $j\to i$. Considering a stimulus then simply corresponds to adding term $\theta_{0i} P_0$ to the sum in \cref{eq_sigma_base}.
This function can be interpreted as a simplification of the mechanistic form used in~\cite{Herbach2017,Bonnaffoux2019} and there is also a similarity with the form used in~\cite{Ocone2013}. Such interactions take various forms in literature, and here the most important feature is the non-Gaussian structure of the resulting \enquote{intrinsic noise}~\cite{Smirnov2018}.

Finally, to reduce the number of parameters and since protein levels $P_i$ are not observed, we set
\begin{equation}\label{eq_s1}
s_{1,i} = \frac{d_{0,i} d_{1,i} \koff{i}}{k_{1,i} s_{0,i}}
\end{equation}
so that each $P_i$ is scaled to span the same range. Importantly, this has no effect on the model dynamics since $\kon{i}$ depends on terms $P_j$ only through a linear combination with parameters $\theta_{ji}$ in~\cref{eq_sigma_base}; it further makes these parameters comparable with each other.

\subsection{Bursty regime}\label{seq_bursty_model}

Compatibly with real single-cell data~\cite{Raj2006,Suter2011,Vinuelas2013,Singer2014,Albayrak2016,Richard2016,Semrau2017}, we consider the so-called bursty regime of the previous model. It corresponds to the experimentally observed situation where active periods are short but characterised by a high transcription rate, thereby generating \enquote{bursts} of mRNA~\cite{Donovan2019,Tantale2016,Chong2014}.
Mathematically speaking, this regime is derived from the original model by taking $\koff{i} \to +\infty$ at the same time as $s_{0,i} \to +\infty$ while keeping $\koff{i}/s_{0,i}$ fixed, for all $i = 1, \dots, n$.

\begin{figure}[t]
\centering
\includegraphics[width=0.91\textwidth]{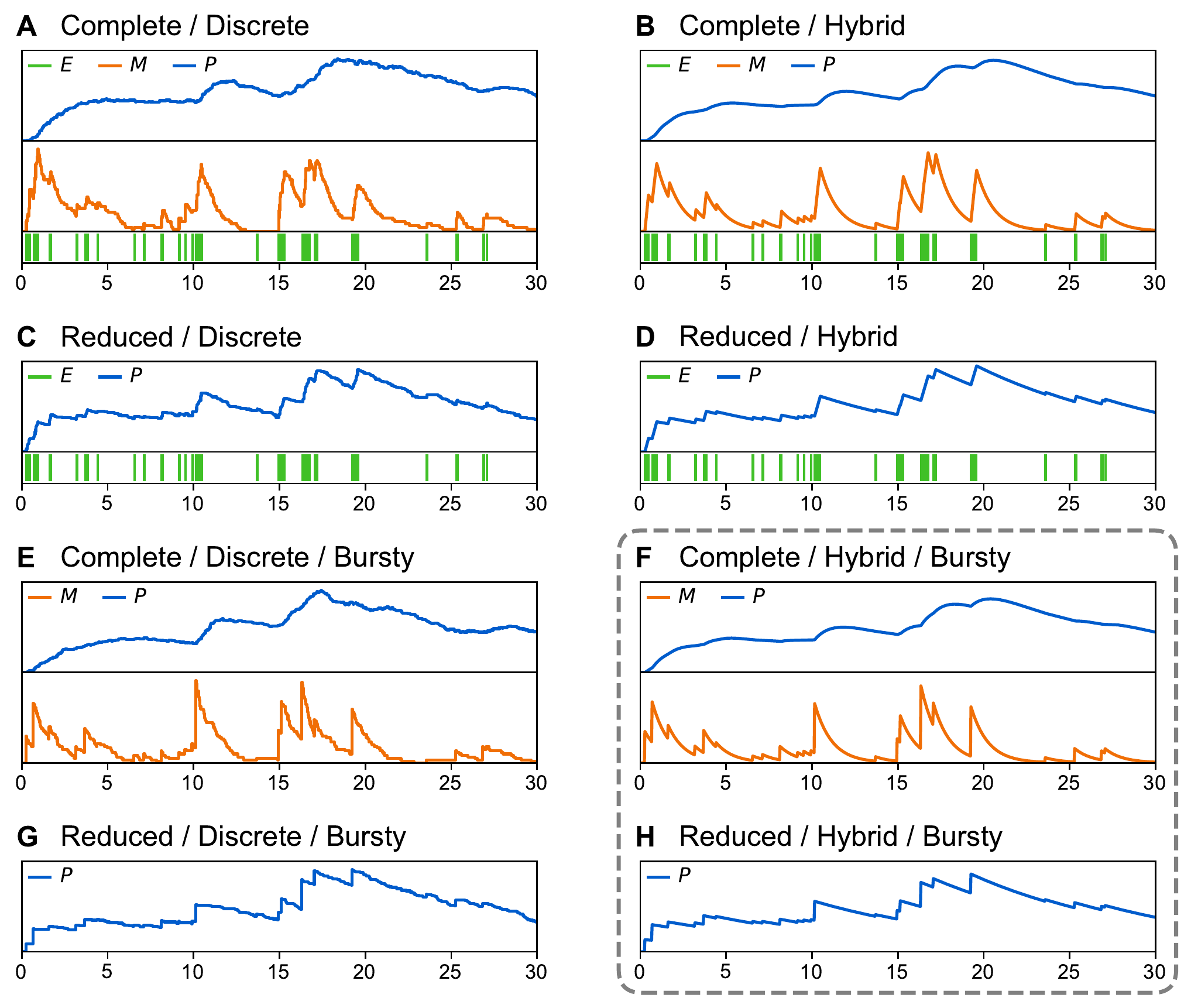}
\caption{Sample time trajectories for various stochastic gene expression models. Promoter activity, mRNA levels and protein levels are denoted by $E$, $M$ and $P$, respectively. These models all derive from a two-state promoter and are characterized by combinations of the following features: complete models (A, B, E, F) describe both mRNA and protein production while reduced models (C, D, G, H) do not describe mRNA; discrete models (A, C, E, G) account for precise molecule numbers while hybrid models (B, D, F, H) combine discrete events with ODEs; bursty models (E, F, G, H) correspond to the limiting case where promoter active periods can be considered as instantaneous regarding the global dynamics. In this work, we use specifically two models indicated by the dashed gray box: the first one (F) as an efficient compromise for simulating gene regulatory networks, and the second one (H) as a proxy to derive the inference procedure.
All trajectories are based on the same parameter values ($\konb = 1$, $\koffb = 10$, $d_0 = 1$, $d_1 = 0.1$, $s_0 = 100$, $s_1 = 1$).}
\label{fig2}
\end{figure}

The promoter states $E_i$ then no longer need to be described since active periods are infinitely short: we still obtain a well-defined PDMP~\cite{Crudu2012}, but random jumps now directly affect the trajectory of $M_i$.
The construction of a trajectory is as follows: starting from state $(M(t),P(t))$, the dynamics is given by
\begin{align}
\dot{M_i} &= - d_{0,i} {M_i} \label{eq_bursty_m} \\
\dot{P_i} &= s_{1,i} {M_i} - d_{1,i} {P_i} \label{eq_bursty_p}
\end{align}
until a burst occurs for gene $i$, still at rate $\kon{i}(P)$, and then $M_i$ jumps by a random height according to exponential distribution with rate parameter $\koff{i}/s_{0,i}$. The dynamics again follows \cref{eq_bursty_m} and \cref{eq_bursty_p} until next burst, and so on: an example of trajectory is given in \cref{fig2}F and the gamma distribution arises as the exact steady-state distribution of $M_i$ when $\kon{i}$ is constant (\cref{tab_models}). As done in~\cite{Kim2013}, we then interpret $s_{0,i}/\koff{i}$ as the average \emph{burst size} and $\kon{i}(P)$ as the \emph{burst frequency} of gene $i$ conditionally to $P$.

\begin{table}[b]
\centering\small
\caption{Known stationary distributions for models of \cref{fig2}, with $M$ and $P$ denoting mRNA and protein levels as before. Here $\Bet(a,b)$, $\Gam(a,b)$ and $\Poi(c)$ denote random variables following the related distributions (with $c$ itself a random variable). The distribution of $P$ in complete models (A, B, E, F) is still lacking in terms of analytical results, which motivates reduced models (C, D, G, H) where $P$ is in fact treated as $M$ except for a change of parameters. Besides, while such explicit representations allow to easily understand in which range of parameters discrete models (left column) are well approximated by their hybrid counterparts (right column), it is also important to note that the two model families happen to be linked analytically, regarding both stationary and time-dependent distributions~\cite{Dattani2017,Herbach2019}.}
\begin{tabular}{@{}llll@{}}
\hline
Distribution & References & Distribution & References \\
\hline \\[-1em]
(A) \; $M \sim \Poi\left(\frac{s_0}{d_0}\cdot\Bet\left(\frac{\konb}{d_0},\frac{\koffb}{d_0}\right)\right)$ & \cite{Kim2013,Dattani2017} & (B) \; $M \sim \frac{s_0}{d_0}\cdot\Bet\left(\frac{\konb}{d_0},\frac{\koffb}{d_0}\right)$ & \cite{Dattani2017,Herbach2019} \\ \\[-0.75em]
(C) \; $P \sim \Poi\left(\frac{s_0s_1}{d_0d_1}\cdot\Bet\left(\frac{\konb}{d_1},\frac{\koffb}{d_1}\right)\right)$ & \cite{Dattani2017,Herbach2017} & (D) \; $P \sim \frac{s_0s_1}{d_0d_1}\cdot\Bet\left(\frac{\konb}{d_1},\frac{\koffb}{d_1}\right)$ & \cite{Dattani2017,Herbach2017} \\ \\[-0.75em]
(E) \; $M \sim \Poi\left(\Gam\left(\frac{\konb}{d_0},\frac{\koffb}{s_0}\right)\right)$ & \cite{Shahrezaei2008,Herbach2017} & (F) \; $M \sim \Gam\left(\frac{\konb}{d_0},\frac{\koffb}{s_0}\right)$ & \cite{Malrieu2015,Herbach2017} \\ \\[-0.75em]
(G) \; $P \sim \Poi\left(\Gam\left(\frac{\konb}{d_1},\frac{d_0\koffb}{s_0s_1}\right)\right)$ &  \cite{Shahrezaei2008,Friedman2006} & (H) \; $P \sim \Gam\left(\frac{\konb}{d_1},\frac{d_0\koffb}{s_0s_1}\right)$ & \cite{Friedman2006,Mackey2011} \\ \\[-1em]
\hline
\end{tabular}
\label{tab_models}
\end{table}

\begin{figure}[t]
\includegraphics[width=\textwidth]{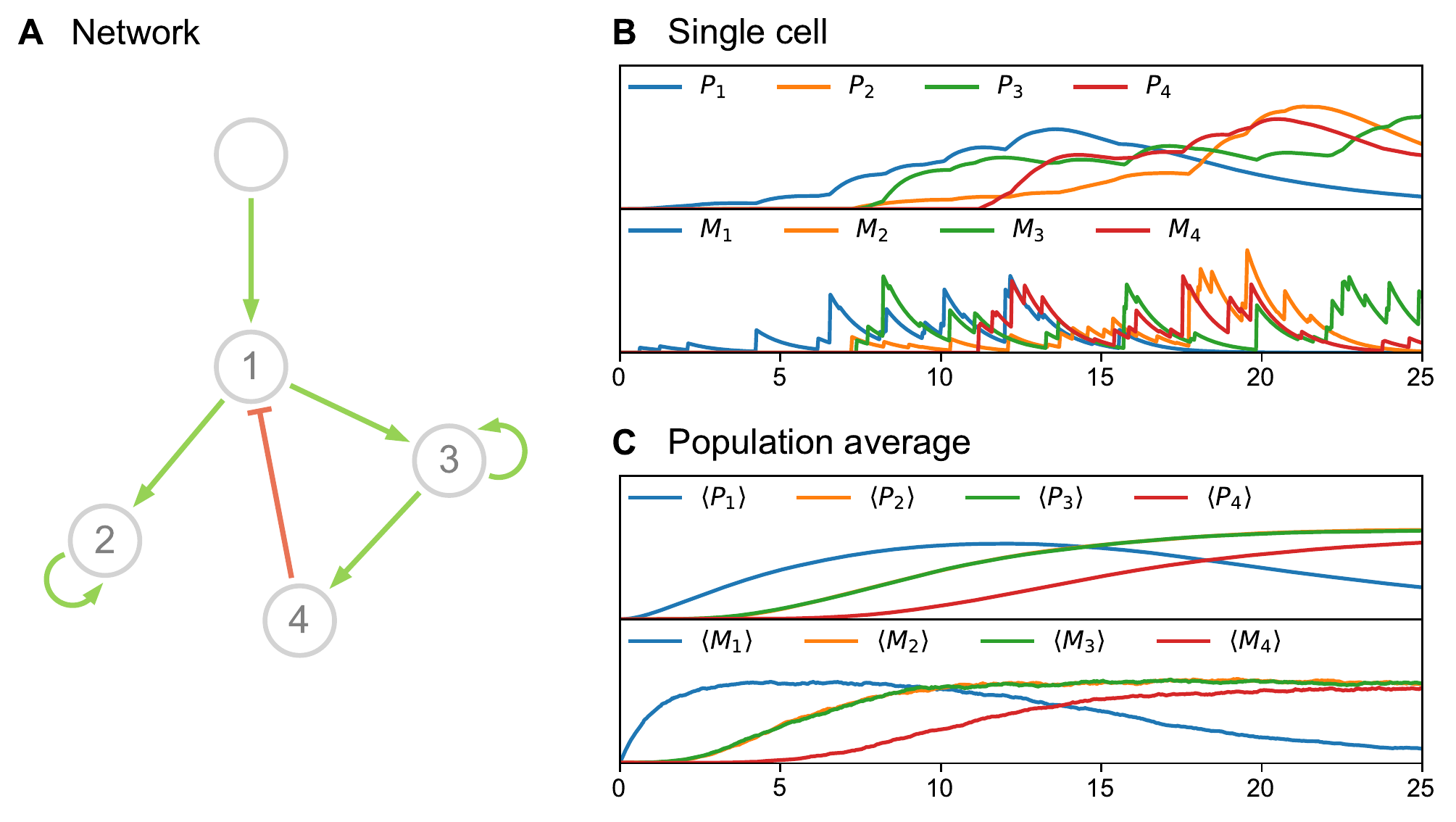}
\caption{Modeling part of the HARISSA package, used for data simulation. (A) Example of 4-gene network with a stimulus. Although such a small network may appear very simple, it already has some interesting features (branching, loop, inhibition) which make it interesting in a first phase of benchmarking (see \cref{fig5}). (B) Simulated trajectory of mRNA and protein levels along time (in hours) for one cell. (C) Population-averaged trajectory (2000 cells). As visible in this example, the notion of \enquote{average cell} is not sufficient to describe the system since individual cell trajectories strongly deviate from the mean.}
\label{fig3}
\end{figure}

This model will be our standard for simulations and benchmarking: contrary to ordinary and stochastic differential equations commonly used to model gene regulatory networks, it directly leads to gamma-like distributions~\cite{Friedman2006,Mackey2011}, a very common feature of single-cell datasets~\cite{Albayrak2016,Singer2014}.
For simplicity, simulations in this article are all based on same $d_{0,i}$ and $d_{1,i}$ values such that for each gene, mRNA and protein half-lives are $\SI{9}{h}$ and $\SI{46}{h}$ respectively~\cite{Schwanhausser2011}.
It should be noted that unlike the hypothesis made in~\cite{Shahrezaei2008}, where mRNA levels stay very small (one or two molecules) and only proteins are considered as bursty, the bursts defined here first come from mRNA levels (thereby spanning a high range), while proteins rather tend to buffer such bursts (\cref{fig2}).

\subsection{Separation of time scales}\label{seq_bursty_prot}

Although the bursty mechanistic model can be efficiently simulated (see \cref{sec_simulation}), it is still too complex to be used for inference. Here we derive final simplifications that lead to a tractable statistical model: this step is detailed in~\cite{Herbach2017} but we briefly recall it here for clarity.

In line with literature~\cite{Schwanhausser2011}, we place ourselves in the situation where the dynamics of mRNA is faster than that of proteins (i.e., $d_{0,i} \gg d_{1,i}$).
By separating these two time scales as done in~\cite{Herbach2017}, we obtain a simplified network model in which only proteins are described: it simply consists in degradation equations
\begin{equation}\label{eq_bursty_p_separate}
\dot{P_i} = - d_{1,i} {P_i}
\end{equation}
plus bursting as described in \cref{seq_bursty_model}, now affecting $P_i$ with frequency $\kon{i}(P)$ and average burst size $(s_{0,i}s_{1,i})/(d_{0,i}\koff{i})$.
We note that several authors also considered this model, with a possible generalization of the burst size distribution and other $\kon{i}(P)$ forms, for a single gene~\cite{Friedman2006,Mackey2011} and more recently for gene networks~\cite{Pajaro2017}.
The related master equation, characterizing the time-dependent distribution of $P$, turns out to be integro-differential and is provided in \cref{seq_master_equation}. The inference algorithm presented in \cref{sec_inference_procedure} is mainly based on the stationary solution of this master equation.

In addition, it is reasonable to assume that promoters are faster than proteins, that is, $k_{1,i} \gg d_{1,i}$. We thereby obtain, along with the above dynamical model for $P$, a quasi-steady-state approximation for the conditional distribution of $M$ given $P$. Under this approximation, mRNA levels $M_i$ are independent conditionally to $P$, each $M_i$ following a gamma distribution characterized by rate $\kon{i}(P)/d_{0,i}$ and shape $\koff{i}/s_{0,i}$~\cite{Herbach2017}.

\subsection{Self-consistent proteomic field}\label{seq_scpf}

This last simplification step is at the core of our approach. While no analytic expression for the stationary distribution of $P$ seems available as an exact solution of the master equation, it is possible to consider instead a self-consistent proteomic field (SCPF) approximation such as presented in~\cite{Walczak2005} and applied, for example, in~\cite{Kim2007,Zhang2014}. The principle is to split the master equation into $n$ independent one-dimensional equations corresponding to each gene $i=1,\dots,n$ by clamping every $P_j$ for $j \neq i$, and then taking the product of their stationary solutions as a proxy for the true distribution~\cite{Herbach2017}.
Interestingly, this is similar to the Hartree approximation in physics and it turns out to be closely related to the notion of pseudo-likelihood in multivariate statistics~\cite{Besag1975,Herbach2018}.

In our case and using the particular scaling \cref{eq_s1}, each term of the product corresponds to the density function of a gamma distribution with rate $c_i\sigma_i(P)$ and shape $c_i$, where $\sigma_i(P)$ is defined as in \cref{eq_sigma_base} and $c_i = k_{1,i}/d_{1,i}$. As a result, it can be mathematically shown that when $k_{1,i} \gg d_{1,i}$, the SCPF  distribution is concentrated on the set of fixed points of $\sigma = (\sigma_1,\dots,\sigma_n)$, that is, points $P$ such that $\sigma_i(P) = P_i$ for all $i=1,\dots,n$~\cite{Herbach2018}.
In order to further simplify inference, we therefore replace $\kon{i}(P)=k_{1,i}\sigma_i(P)$ with $k_{1,i}P_i$ in the conditional distribution of $M_i$ given $P$.

Finally, since our aim is to use count data such as single-cell RNA-Seq, we make the assumption~\cite{Sarkar2021} that the data for gene $i$ at time $t$ is sampled from the Poisson distribution with parameter $M_i(t)$, each cell corresponding to one copy of the PDMP model. Hence, conditionally to $P$, we do not consider gamma distributions but rather gamma-Poisson, that is, negative binomial distributions.
Note that when each $\kon{i}$ is constant, the same negative binomial distributions can be derived directly from the bursty regime of the original discrete-value model~\cite{Shahrezaei2008}.

\section{Inference procedure}\label{sec_inference_procedure}

As only mRNA levels are observed, the main idea of our inference strategy consists of treating protein levels as latent variables and using the above approximation of mRNA distribution as a statistical likelihood for the data.
In practice, although it is difficult to assess in a comprehensive way, the approximation accuracy seems satisfactory and inference indeed corresponds to calibration of the mechanistic model, thereby providing a high level of biological interpretability~\cite{Herbach2017}.
Assuming that cells are independent regarding their gene expression, we build upon this approach and present a relevant procedure for performing network inference from single-cell data.

\subsection{Statistical framework}

As mentioned in the introduction, it is unlikely that all interactions can be observed at each time point: this is precisely one of the reasons for collecting time-stamped data corresponding to the evolution of the system after applying a perturbation such as cell medium change.
Contrary to what was done in~\cite{Herbach2017}, we shall distinguish the \emph{mechanistic} network parameter $\theta = (\theta_{ij})$ denoting biological interactions $i \to j$ (network structure) and a time-dependent \emph{descriptive} network parameter $\alpha(t) = (\alpha_{ij}(t))$ representing statistical dependencies $i \to j$ that are detectable in data at time point $t$ (network states).
Intuitively, a nonzero value for $\theta_{ij}$ will result in nonzero values of $\alpha_{ij}(t)$ at only some time points $t$.

The resulting likelihood function is detailed in \ref{seq_likelihood}. Basically, it is derived from the approximate distributions of mRNA and protein levels introduced in \cref{seq_scpf} by simply replacing $\theta$ with $\alpha(t)$ after reintroducing a time variable $t$.
Although identifiability of $\alpha(t)$ cannot be guaranteed, the idea of an underlying information flux---starting from the stimulus and then progressing throughout the network---suggests that one may recover all active parts of the network structure, including feedback loops, as long as time points are sampled with sufficient precision~\cite{Bonnaffoux2019}.

\begin{figure}[t]
\centering
\includegraphics[width=\textwidth]{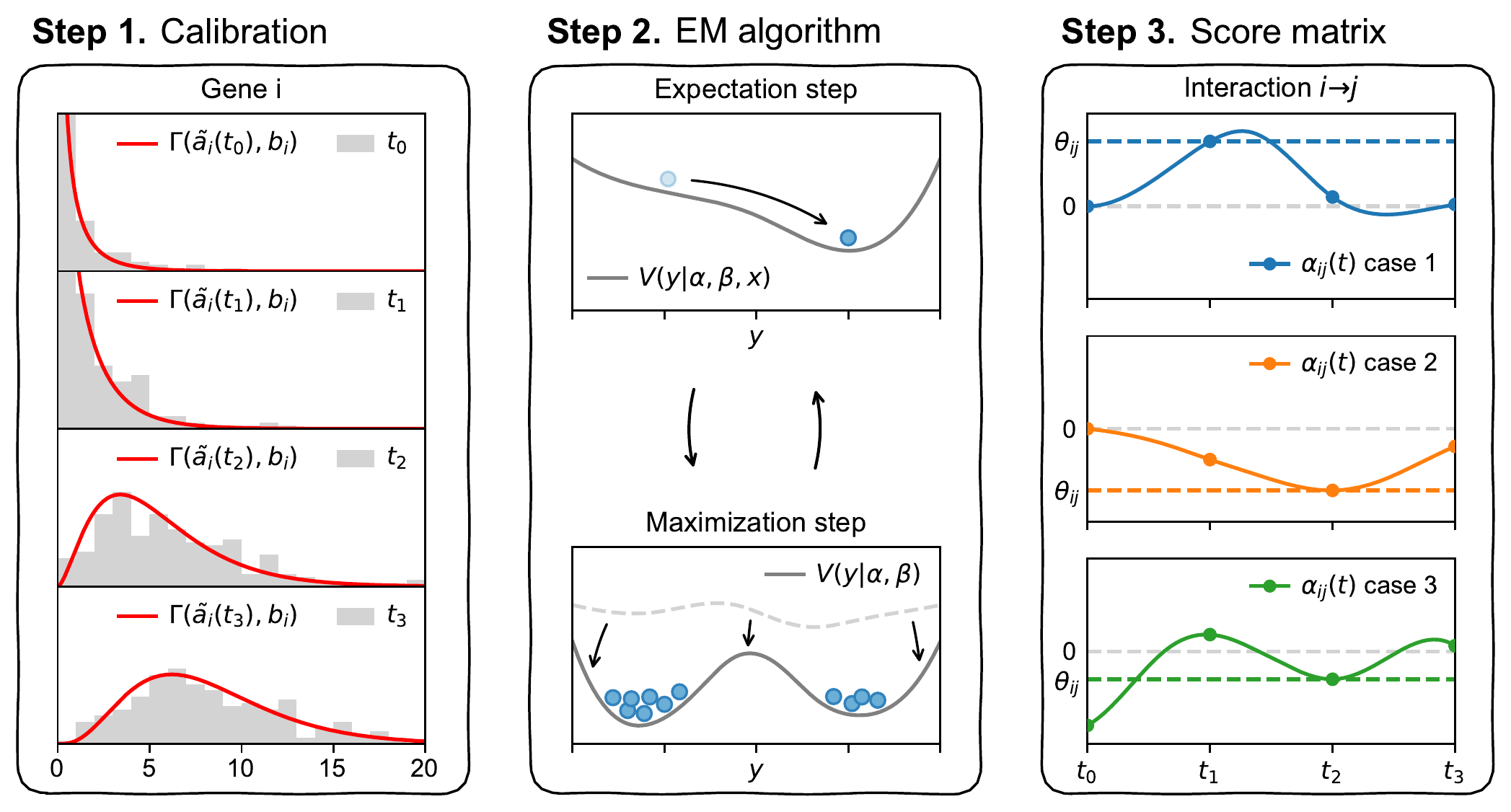}
\caption{Inference module of the HARISSA package, consisting of 3 distinct steps. (Step 1) For each gene $i$, kinetic parameters $a_i$ and $b_i$ are estimated using the related marginal distribution for all time points (see \cref{seq_gene_calibration}), with $a_i = \max_t\{\tilde{a}_i(t)\}$ corresponding to the bursting frequency of gene $i$ when it is fully expressed. (Step 2) Network parameters $\alpha$ and $\beta$ are estimated using an EM algorithm, that is, alternately performing expectation (E) and maximization (M) steps until convergence (see \cref{seq_network_parameters}). As part of the self-consistent proteomic field approximation, the E step can be interpreted as finding, for each cell $k$, protein levels $\mathbf{y}_k = (y_{ki})$ that minimize energy potential $V(\mathbf{y}_k | \alpha, \beta, \mathbf{x}_k)$ given the current values of $\alpha$ and $\beta$ as well as observed mRNA levels $\mathbf{x}_k = (x_{ki})$. The M step can then be interpreted as updating $\alpha$ and $\beta$ in order to minimize the average energy $\langle V(\mathbf{y} | \alpha, \beta) \rangle$ of protein levels in all observed cells. (Step 3) The final interaction matrix $\theta = (\theta_{ij})$ is obtained from time-dependent descriptive parameter $\alpha = (\alpha_{ij}(t))$ using the formula $\theta_{ij} = \alpha_{ij}(t^*)$ where $t^* = \arg\max_{t>0}\{|\alpha_{ij}(t)|\}$, that is, time $t>0$ at which interaction $i\to j$ is strongest.}
\label{fig4}
\end{figure}

\subsection{Gene calibration}
\label{seq_gene_calibration}

Thanks to the time scale separation introduced in \cref{seq_bursty_prot} and assuming $k_{0,i}=0$, we only have three aggregated kinetic parameters defined by
\begin{equation}\label{eq_knetic_param}
a_i = \frac{k_{1,i}}{d_{0,i}} , \quad b_i = \frac{\koff{i}}{s_{0,i}} , \quad c_i = \frac{k_{1,i}}{d_{1,i}}
\end{equation}
with $a_i$ and $b_i$ being identifiable from the distribution of $M_i$, and $c_i$ controlling the \enquote{noise amount} of $P_i$.
In this article, we estimate $a_i$ and $b_i$ using the same preprocessing algorithm as in~\cite{Herbach2017}, which can be seen as a basic calibration step for each gene, and we set $c_i = 10$ for all $i=1,\dots,n$ as a default value in the absence of additional data, in line with quantitative results comparing mRNA and protein kinetics~\cite{Schwanhausser2011}. Unlike~\cite{Herbach2017}, there is no other kinetic parameter: the only remaining parameters, which we shall refer to as network parameters, are interactions $\theta_{ij}$, to be reconstructed from $\alpha_{ij}(t)$, and basal activity $\beta_i$.

As an additional effect, this preprocessing step also discards genes whose variations are too small to be detected and which therefore appear unaffected by the initial perturbation: this is entirely relevant since such genes are unlikely to belong to the part of the network that is active during the experiment.

\subsection{Network parameters}
\label{seq_network_parameters}

After the above preprocessing steps, we infer $\beta = (\beta_i)$ and $\alpha(t) = (\alpha_{ij}(t))$ using the same classification EM algorithm as detailed in~\cite{Herbach2017}, that is, alternately maximizing the likelihood with respect to these parameters and with respect to latent protein levels. Here we add the constraint that $\alpha_{ij}(t) = 0$ for every interaction $i \to j$ that does not belong to the preselection. In addition, self-regulation is not considered (i.e., $\alpha_{ii}(t) = 0$ for all $i=1,\dots,n$ and for all $t$) as it is notoriously difficult to infer reliably within this framework.

Once $\alpha(t)$ has been inferred for all time points $t$, we construct the mechanistic network parameter $\theta$ as follows: for each interaction $i\to j$, we set $\theta_{ij} = \alpha_{ij}(t)$ where $t$ is the positive time point for which $\alpha_{ij}(t)$ has maximum absolute value, ignoring dependencies that are only detected before stimulus.
Note that the resulting \enquote{non-Markov} paradigm for mRNA can be interpreted more precisely in terms of a hidden Markov model~\cite{Stumpf2017} where proteins play the role of the latent Markov structure.

\section{Results}\label{sec_results}

In this section, we first evaluate the performance of HARISSA in inference---defined here simply as the correctness of $\theta$ nonzero entries---by simulating data corresponding to various networks, and then illustrate its practical application using real data from a differentiation experiment~\cite{Semrau2017}.

\subsection{Small network}

\begin{figure}[t]
\includegraphics[width=\textwidth]{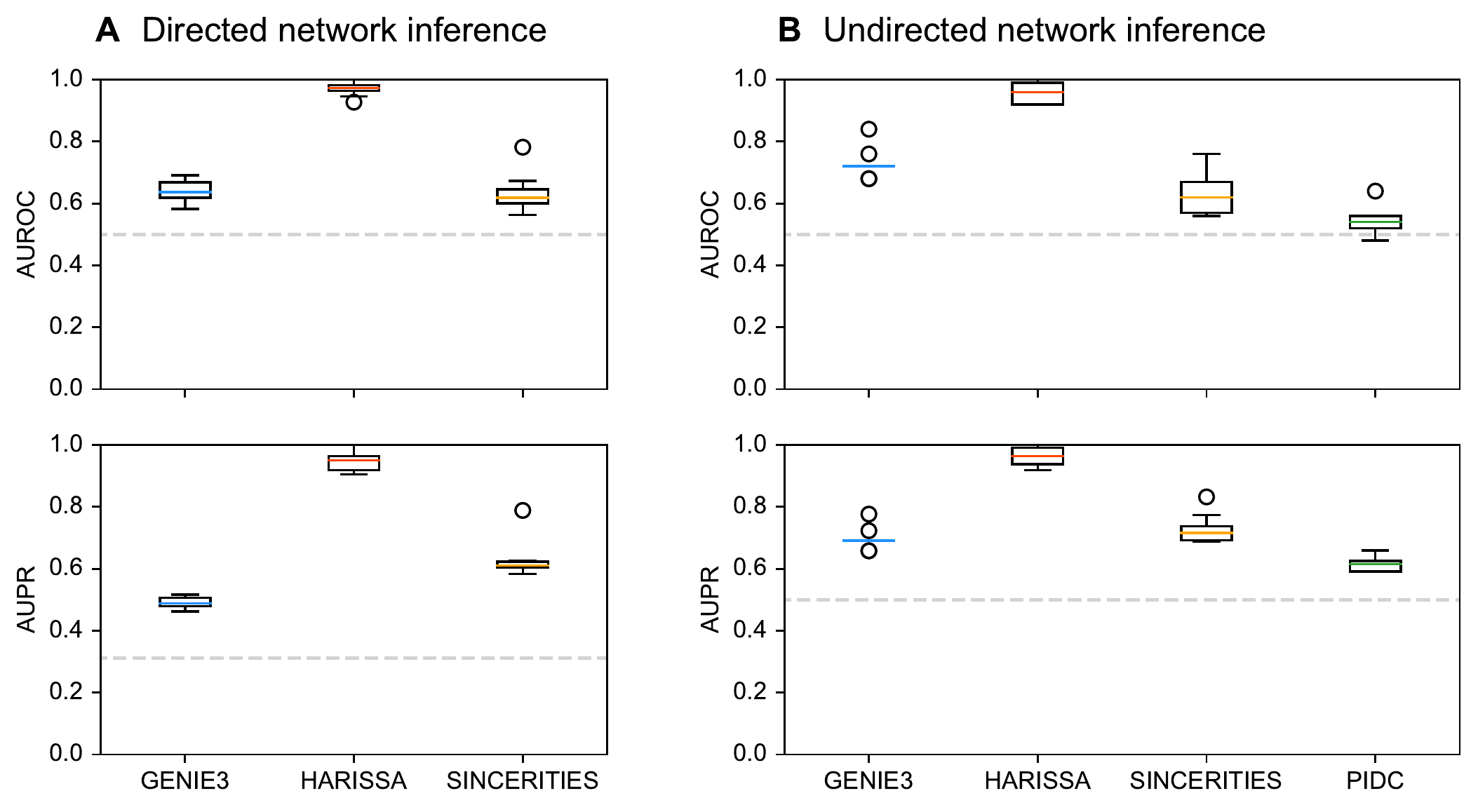}
\caption{Inference results for the 4-gene network of \cref{fig3}. Performances of GENIE3, HARISSA, SINCERITIES and PIDC are measured in terms of areas under receiver operating characteristic curves (AUROC) and areas under precision-recall curves (AUPR) obtained for 10 independently simulated datasets. Each dataset contained the same 10 time points and $50$ cells per time point (500 cells sampled per dataset). Only HARISSA and SINCERITIES exploited the time information while GENIE3 and PIDC used merged time points. The dashed gray line indicates the average score that would be obtained by the random estimator (detecting a link or not with equal probability). (A) Directed network description with unsigned interactions (PIDC being unable to infer directions). (B) Undirected network description with unsigned interactions.}
\label{fig5}
\end{figure}

We first tested the performance of HARISSA on the four-gene network of \cref{fig1}. To this aim, we simulated several datasets by sampling independent cells at 10 time points $t = 0$, $2$, $4$, $6$, $8$, $11$, $13$, $15$, $17$, and $\SI{20}{h}$ (same time scale as \cref{fig2}) with 50 cells per time point (500 cells in total per dataset). Inference was then independently performed for ten such datasets and the results were merged into receiver operating characteristic (ROC) and precision-recall (PR) curves whose areas are shown in \cref{fig5}.

\begin{figure}[t]
\includegraphics[width=\textwidth]{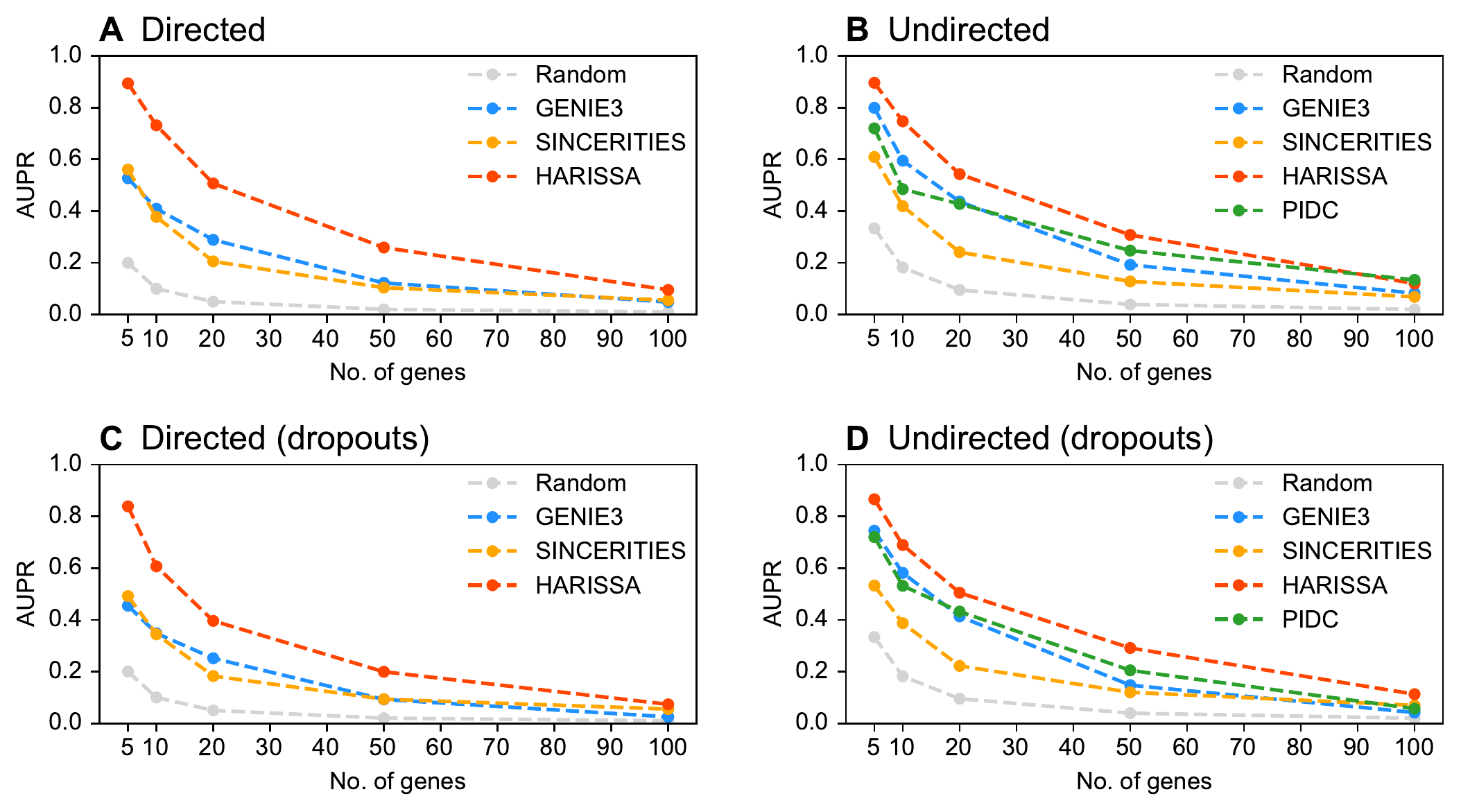}
\caption{Inference results for tree-like networks with different numbers of genes. Performances of GENIE3, HARISSA, SINCERITIES and PIDC are measured in terms of area under precision-recall curve (AUPR), each point representing an average value based on 10 datasets corresponding to ten random tree-like activation networks of equal size. Each dataset contained the same 10 time points and $100$ cells per time point (1000 cells sampled per dataset). The dashed gray line represents the average score that would be obtained by the random estimator (detecting a link or not with equal probability). (A, C) Directed network description with unsigned interactions (PIDC being unable to infer directions). (B, D) Undirected network description with unsigned interactions. (C, D) Inference results in the case of data being affected by technical dropouts, thus losing information. Given a certain dropout rate, all methods seem to deteriorate similarly.}
\label{fig6}
\end{figure}

Note that here PR may be a better performance indicator than ROC because of the well-known \emph{class imbalance} problem. Indeed, as stated in the previous section, the number of actual interactions is expected to be smaller than half of the total number $n\times(n+1)$ of possible interactions (e.g., the ratio is $5/16 \approx 0.31$ in the network of \cref{fig1}), making $\theta_{ij} \neq 0$ (positives) not comparable with $\theta_{ij} = 0$ (negatives).
In line with this observation, we shall focus on avoiding false positives (interactions that are detected but not actually present) rather than false negatives (interactions that are present but not detected), which is reflected by the preference of PR over ROC.

We compared HARISSA with two other methods, GENIE3~\cite{Huynh-Thu2010} as a widely used standard algorithm for gene network inference, and SINCERITIES~\cite{Papili-Gao2018} which is one of the few available methods to exploit the time information. This algorithm first computes individual variations of each gene, defined as the sequence of Kolmogorov--Smirnov distances between marginal distributions related to consecutive time points, and then performs network inference as a set of penalized linear regressions, assuming that the variation of a gene at a given time point linearly depends on the variations of other genes at the previous time point.

As shown in \cref{fig5}, the HARISSA procedure turned out to perform better than the other two in terms of both ROC and PR curves.
More generally, the aim is to perform at least as well as established algorithms such as GENIE3, while providing biological interpretability---thanks to the associated mechanistic model---and higher precision (proportion of true interactions among those detected).
In this example, since genes 2 and 3 have the same input (gene 1), their marginal distributions evolve equally at each time point (see \cref{fig3}B) so SINCERITIES alone is unable to distinguish them, illustrating the improvement that is typically expected from HARISSA.

\subsection{Larger networks}

We then considered three cases of tree-like activation networks ($5$, $10$, $20$, $50$, and $100$ genes). Each case was based on simulating ten datasets corresponding to ten random networks of the same size, sampled from the uniform distribution over trees rooted in the stimulus. All datasets contain the same 10 time points $t = 0$, $2$, $5$, $8$, $11$, $13$, $16$, $19$, $22$, and $\SI{25}{h}$, with $100$ cells per time point (1000 cells in total per dataset).
As before, inference was then independently performed for all datasets using GENIE3, HARISSA and SINCERITIES: the results, measured in terms of area under ROC (AUROC) and area under PR (AUPR), are shown in \cref{fig6}.

Unsurprisingly, the choice of time points turned out to be crucial: while sufficient cells per time point must be sampled, a sequence of time points that is too coarse compared to the dynamics would directly lead to a lack of inference accuracy.
Moreover, because of the delay induced by each intermediate protein, some genes were never activated before $\SI{25}{h}$ (the performance on 50-gene and 100-gene networks could thus be improved by simply adding later time points). This observation supports the idea that when needing to perform some biological function within a given time, the depth of real networks is inherently limited by the half-lives of involved proteins~\cite{Bonnaffoux2019}.

\begin{table}[h]
\centering\small
\caption{Computational times comparison among PIDC, SINCERITIES, GENIE3 and HARISSA. For different sizes, an average runtime is obtained after inferring, for each given size, ten tree-like networks as in \cref{fig6}.
All timings were measured on an 16-GB RAM, $3.1$ GHz Intel Core i7 computer.}
\begin{tabular}{@{}lccrr@{}}
\hline
Runtime & PIDC & SINCERITIES & GENIE3 & HARISSA \\
\hline
5 genes & $\SI{0.02}{s}$ & $\SI{0.03}{s}$ & $\SI{6.67}{s}$ & $\SI{0.29}{s}$ \\
10 genes & $\SI{0.04}{s}$ & $\SI{0.07}{s}$ & $\SI{14.19}{s}$ & $\SI{0.36}{s}$ \\
20 genes & $\SI{0.06}{s}$ & $\SI{0.19}{s}$ & $\SI{28.10}{s}$ & $\SI{0.61}{s}$ \\
50 genes & $\SI{0.18}{s}$ & $\SI{0.76}{s}$ & $\SI{80.76}{s}$ & $\SI{1.28}{s}$ \\
100 genes & $\SI{0.75}{s}$ & $\SI{2.68}{s}$ & $\SI{150.98}{s}$ & $\SI{3.09}{s}$ \\
\hline
\end{tabular}
\label{tab_runtimes}
\end{table}

\subsection{Real data}

Finally, we applied HARISSA to real data from a differentiation experiment on mouse embryonic stem cells~\cite{Semrau2017}. This single-cell RNA-Seq dataset consists of $9$ time points ($t=0$, $6$, $12$, $24$, $36$, $48$, $60$, $72$, and $\SI{96}{h}$) and has the advantage of containing a relatively large number of sampled cells, between $137$ and $335$ cells per time point ($272$ on average).
However, the number of genes ($17452$) is much larger than in the previous examples: for simplicity, we reduce our analysis here to a subset of $41$ preselected genes~\cite{Semrau2017}.
While a thorough analysis is beyond the scope of this article, requiring great caution as the number of cells would become critical, we point out that the computation time of HARISSA is significantly shorter than that of GENIE3 and would remain acceptable with a much larger subset of genes (e.g., those detected as actually varying along time points).

\begin{figure}[t]
\centering
\includegraphics[width=\textwidth]{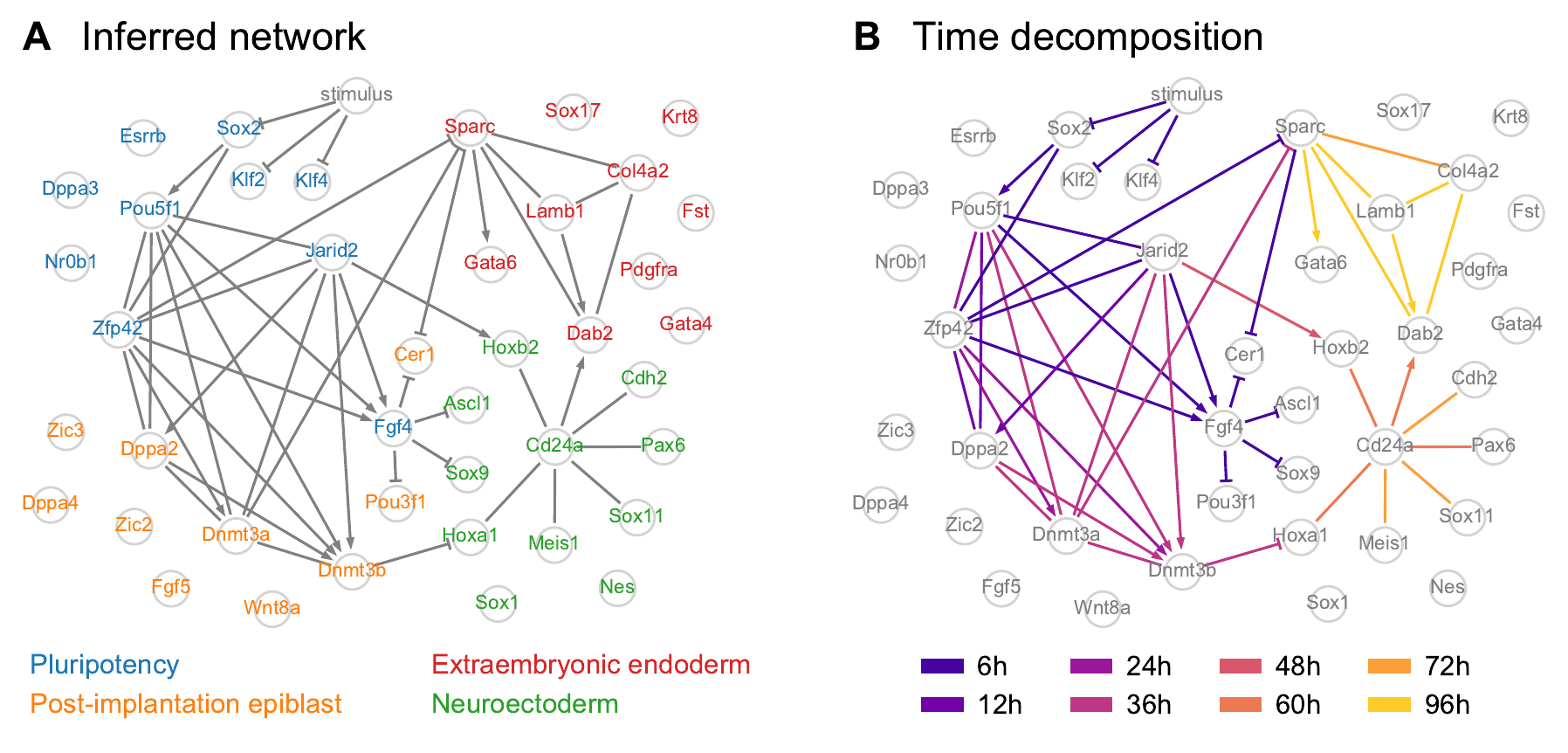}
\caption{Interactions inferred by HARISSA from the single-cell RNA-Seq dataset of~\cite{Semrau2017} using a preselected subset of $41$ genes. (A) Inferred network topology, obtained by retaining only the $4\%$ strongest activations and the $4\%$ strongest inhibitions. The links with no arrowhead represent activations for which both directions were retained. For interpretability purposes, we recall the four gene clusters described in~\cite{Semrau2017}. (B) Underlying time decomposition of inferred interactions. Each color corresponds to an experimental snapshot time and indicates the point at which an interaction was strongest and thus actually inferred (see \cref{fig5}).}
\label{fig7}
\end{figure}

\begin{figure}[t]
\centering
\includegraphics[width=\textwidth]{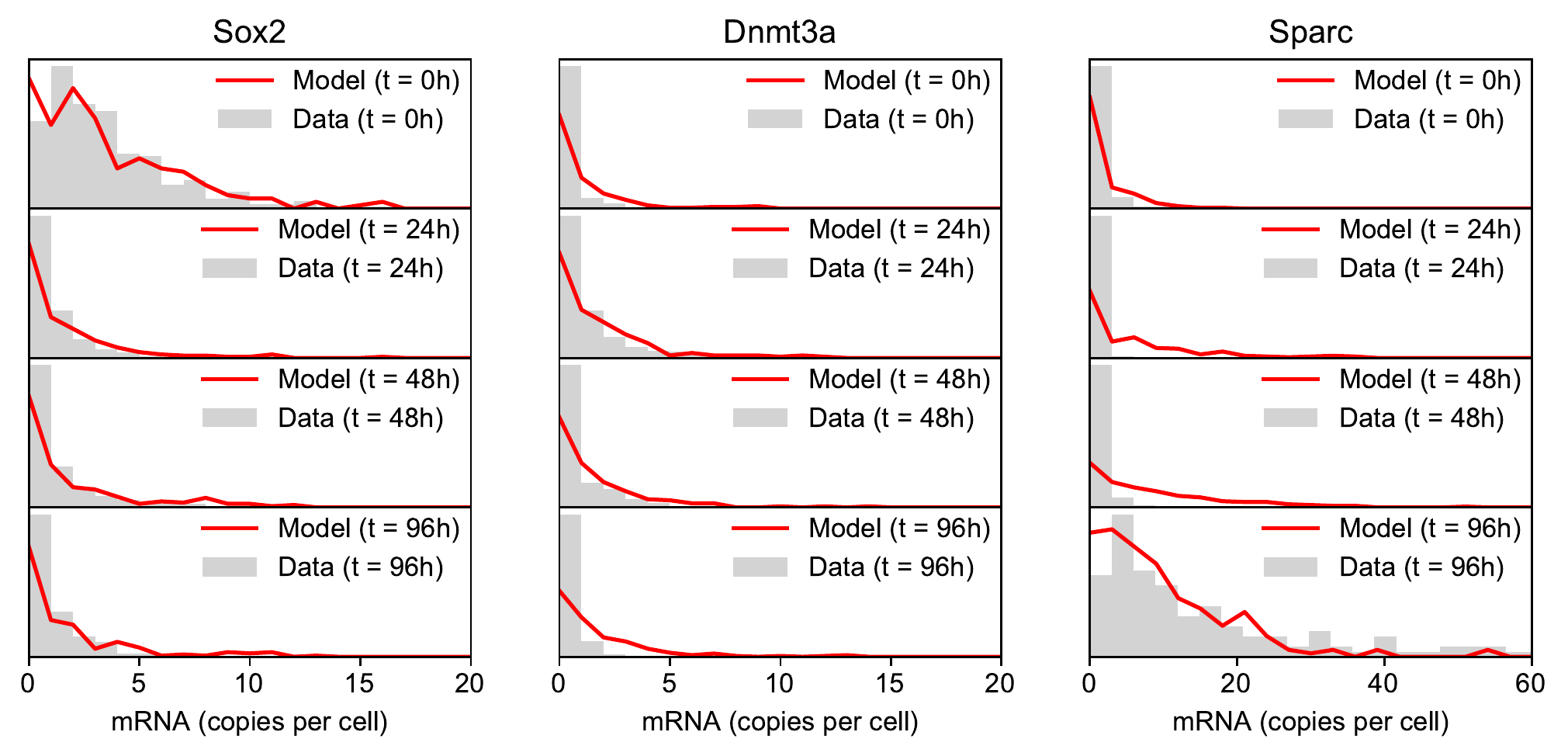}
\caption{Back to the mechanistic model. Starting from the same network as in \cref{fig7} but now considering quantitative parameter values, the mechanistic model was used to simulate new snapshot data (200 cells per time point) and then compared to the original data. Although significant discrepancies persist, the overall dynamics are rather satisfactorily reproduced, supporting the idea that the model has indeed been calibrated through the inference procedure described in \cref{sec_inference_procedure}.}
\label{fig8}
\end{figure}

It is worth noticing that HARISSA -- and most importantly the mechanistic model once calibrated (\cref{fig8}) -- can indeed satisfactorily reproduce the data, in which marginal distributions are close to negative binomial distributions or mixtures of such distributions.
The result of inference is shown in \cref{fig7}: interestingly, the number of nonzero entries of $\alpha(t)$, corresponding to detectable interactions, is maximal at time $t=\SI{6}{h}$ and decreases immediately afterwards, with no evolution after $t=\SI{12}{h}$ except for what appear to be statistical fluctuations.
Overall, this pattern is compatible with the idea that biological interactions are only sequentially detectable, according to the progression of the information initially given by a particular stimulus (\cref{fig3}).

\section{Discussion and prospects}\label{sec_discussion}

In this work, we introduced a simplified version of a previously described mechanistic model of gene regulatory networks~\cite{Herbach2017} in the form of a hybrid, piecewise-deterministic Markov process (PDMP) involving mRNA and protein levels.
We then described the related self-consistent proteomic field (SCPF) approximation and used it to develop an inference algorithm.
These two aspects---exact simulation of the mechanistic model and network inference---were implemented as a Python package called HARISSA.

Our approach is therefore based on explicitly reconstructing the molecular dynamics of mRNA and proteins, providing biological interpretability and, most of all, a systematic way to make quantitative predictions that can then be tested in vitro.
In particular, while the network structure is assumed fixed, the network state can vary over time, which is reflected by distinguishing time points in the inference algorithm (\cref{fig4}).

More specifically, our aim here was to apply HARISSA to real single-cell RNA-Seq data. Various technical problems are imputed to this type of data, especially dropouts (non-biological zeros) and artificial scaling factors induced by sequencing: in first analysis, our method appears structurally robust regarding both dropouts and scaling factors.
Once fitted, our mechanistic model can satisfactorily reproduce the dataset of \cite{Semrau2017} with no need for any external \enquote{noise}. Note that contrary to technical dropouts, transcriptional bursting is a biological phenomenon.

Statistically, we focused on reducing the false positive rate as much as possible, potentially at the expense of a higher false negative rate. This reflects the philosophy that a sparse network with reliable edges is, although potentially incomplete, more biologically informative than a dense network with unreliable edges. Under this criterion, our approach seems to perform better than previously assessed methods~\cite{Chen2018,Pratapa2020} such as GENIE3 (figures \ref{fig5} and \ref{fig6}).

\subsection{On bursty gene expression}

Gene expression is now well acknowledged to be bursty, both at mRNA and protein levels in many eukaryotic cells, and this aspect should be taken into account when dealing with single-cell data and gene regulatory networks.
We point out that our paradigm is very different from the hypothesis made by several authors when modeling bursts~\cite{Shahrezaei2008,Elgart2011,Pajaro2017}. Indeed, these authors assume that mRNA molecules are only present in very small numbers, which is contradicted by typical single-cell experiments using various methods~\cite{Richard2016,Albayrak2016,Semrau2017}.
Our model first describes bursts of mRNA, which then generate bursts of proteins that can actually be buffered depending on degradation rates.

While leading to a simplified protein-only model that is similar to~\cite{Pajaro2017} after separating mRNA and protein time scales~\cite{Herbach2017}, our underlying hypothesis therefore does not concern the range of mRNA levels, which can extend in fact to hundreds of molecules~\cite{Richard2016,Albayrak2016,Schwanhausser2011}.
Importantly, this regime is strongly related to so-called \enquote{refractory} promoters, that is, consisting of one transcriptionally active state and one or more transcriptionally inactive states~\cite{Herbach2019}.
In practice, the observed Gamma or negative binomial distributions can be directly derived from the bursty regime of the simple two-state promoter, which appears as a good compromise between realism and parsimony~\cite{Herbach2017,Herbach2019}.

Besides, the question of transcriptional activity transmission along cell lineages is currently an active research field: although it is still far from being completely solved, we note that our mechanistic model seems compatible with recent results~\cite{Phillips2019}.

\subsection{Joint distribution}

By providing the joint distribution of expression levels, single-cell data naturally contains a co-expression structure without requiring any artificial perturbation such as knockouts. This structure turns out to contain information, but using it to improve our biological understanding of fundamental notions such as cell types or states currently remains challenging~\cite{Crow2018}.
Contrary to testing the presence of each interaction independently, for example using relevant triplets of genes~\cite{Chan2017}, our approach aims at estimating all interactions at once from the full joint distribution.

Besides, adding temporal constraints on the descriptive parameter $\alpha(t)$ could reduce the number of parameters to be inferred and substantially improve inference robustness, for example by implementing the notion of information propagation introduced in~\cite{Bonnaffoux2019}.
It would be especially useful, in this view, to systematically measure mRNA and protein degradation rates in order to have a quantitative overview of the delays that are inherently associated with network topologies.

\subsection{Link with machine learning approaches}

Although the inference part of HARISSA exploits the whole joint distribution of mRNA levels, the way it is performed---alternately maximizing the likelihood with respect to protein levels and with respect to network parameters---makes each iteration of the algorithm fairly close to parallel regressions of each gene on the others. In other words, we independently consider each gene as a \enquote{target} and infer its \enquote{regulators}, similarly to what is often done in descriptive methods such as GENIE3~\cite{Huynh-Thu2010}.

Besides, our choice of sigmoid-based interaction function \cref{eq_sigma_base} makes the approximate likelihood similar to an artificial neural network regarding the latent layer of protein levels (see \cref{seq_likelihood}), and more specifically since the likelihood is maximized with respect to protein levels instead of performing integration. In this view, the latent protein layer does not only have a biological interpretation, but is also a way to predict gene expression profiles without simulating the mechanistic model.

While machine learning procedures are powerful tools to extract information from high-dimensional data such as single-cell RNA-Seq~\cite{Luecken2019}, mechanistic models are needed to understand the biological processes involved.
The network inference method presented here, based on a self-consistent field approximation, represents a step in merging these two viewpoints: as recently emphasized in~\cite{Baker2018}, this type of approach could be very fruitful to exploit single-cell data in general.


\appendix

\section{Exact simulation}\label{sec_simulation}

In order to perform simulations of the mechanistic model such as in \cref{fig2}F, we use an efficient thinning method similarly to what is done in~\cite{Benaim2015}.
Basically, instead of discretizing time and using a Euler scheme as in~\cite{Herbach2017}, the idea consists of sampling jump times with maximum rate $k_{1,1} + \cdots + k_{1,n}$, and then deciding with appropriate rule which ones are actual jumps and which ones are invisible \enquote{phantom jumps}. A great advantage of this method is that it is guaranteed to be exact without requiring any numerical integration, contrary to the basic algorithm~\cite{Malrieu2015}. In practice, this approach seems very appropriate to our context as phantom jumps typically represent a moderate fraction of computed jumps.

\section{Master equation for the bursty regime}\label{seq_master_equation}

For $i = 1,\dots, n$, let $e_i = (0,\dots,0,1,0,\dots,0)$ where the nonzero entry is at index $i$.
The master equation corresponding to the simplified network model presented in \cref{seq_bursty_prot}, describing only proteins, is given by
\begin{equation}
\dr_t u(t,y) = \sum_{i=1}^n [\mathcal{D}_iu(t,y) + \mathcal{B}_iu(t,y)]
\end{equation}
where $u(t,y)$ denotes the time-dependent distribution of protein levels $y = (y_1,\dots,y_n)$, $\mathcal{D}_i$ is the deterministic \emph{degradation} operator defined by
\begin{equation*}
\mathcal{D}_iu(t,y) = \dr_{y_i}[d_{1,i}y_i u(t,y)]
\end{equation*}
and $\mathcal{B}_i$ is the stochastic \emph{bursting} operator defined by
\begin{equation*}
\mathcal{B}_iu(t,y) = \left[\int_{0}^{y_i} \kon{i}(y-he_i) u(t,y-he_i) c_i e^{-c_i h} \intd{h}\right] - \kon{i}(y) u(t,y)
\end{equation*}
with $\kon{i}$ defined as in \cref{seq_bursty_model} and $c_i = k_{1,i}/d_{1,i}$.

\section{Likelihood function}\label{seq_likelihood}

Let $n$ still denote the number of genes and $m$ be the number of cells. Each cell $k = 1, \dots, m$ is associated with a particular observation time point $t_k$.
In line with \cref{sec_inference_procedure}, we introduce the following notation:
\begin{itemize}
\item $\mathbf{x}_k = (x_{ki})\in\{0,1,2,\dots\}^{n}$ observed mRNA levels (cell $k$, gene $i$);
\item $\mathbf{y}_k = (y_{ki})\in(0,+\infty)^{n}$ latent protein levels (cell $k$, gene $i$);
\item $\alpha = (\alpha_{ij}(t_k))\in\R^{n \times n}$ interaction $i \to j$ at time $t_k$;
\item $\beta = (\beta_i)\in\R^{n}$ basal activity of gene $i$.
\end{itemize}
The stimulus is represented as gene $i=0$ and we therefore add parameters $\alpha_{0j}(t_k)$ for $j=1, \dots, n$ and $k=1, \dots, m$. We further set $y_{k0} = 0$ if $t_k \leqslant 0$ (before perturbation) and $y_{k0} = 1$ if $t_k > 0$ (after perturbation).
The statistical model is then defined by
\begin{equation}
p(\mathbf{y}_k | \alpha, \beta) = \prod_{i=1}^n {y_{ki}}^{c_i \sigma_{ki} - 1} e^{-c_i y_{ki}} \frac{{c_i}^{c_i\sigma_{ki}}}{\Gamma(c_i\sigma_{ki})}
\end{equation}
and
\begin{equation}
p(\mathbf{x}_k | \mathbf{y}_k) = \prod_{i=1}^n \frac{1}{x_{ki}!} \frac{\Gamma(a_i y_{ki} + x_{ki})}{\Gamma(a_i y_{ki})} \frac{{b_i}^{a_i y_{ki}}}{(b_i+1)^{a_i y_{ki} + x_{ki}}}
\end{equation}
where
\begin{equation}\label{eq_sigma_stat}
\sigma_{ki} = \frac{\exp(\beta_i + \sum_{j} \alpha_{ji}(t_k) y_{kj})}{1 + \exp(\beta_i + \sum_{j} \alpha_{ji}(t_k) y_{kj})}
\end{equation}
with kinetic parameters $a_i$, $b_i$ and $c_i$ being defined as in \cref{eq_knetic_param}, and the sum in \cref{eq_sigma_stat} being over $j=0,1,\dots,n$ to account for the stimulus term $\alpha_{0i}(t_k) y_{k0}$.

Note that $p(\mathbf{y}_k | \alpha, \beta)$ is in general only a pseudo-likelihood as $\sigma_{ki}$ depends on $\mathbf{y}_k$. It happens to be a genuine likelihood when the underlying graph of $\alpha(t)$ is a directed acyclic graph (DAG) for all $t$, but even in this case it is only an approximate solution of the master equation. Finally, we add a standard ridge penalization to each parameter $\alpha_{ij}(t)$.

\phantomsection
\pdfbookmark[section]{Code availability}{availability}
\section*{Code availability}

The dynamical model and the inference procedure were implemented together as a Python package called HARISSA available at \url{https://github.com/ulysseherbach/harissa}.


\phantomsection
\pdfbookmark[section]{References}{references}
\pagestyle{plain}
\setstretch{1}
\small

\bibliographystyle{ieeetr}

\end{document}